\documentclass[10pt,letterpaper,english]{article}
\usepackage{babel}
\usepackage[dvips]{graphics}
\usepackage{epsfig}
\usepackage{latexsym}

\usepackage{amscd,amsmath,amssymb}

\numberwithin{equation}{section}
\newcommand{\D}[1]{\operatorname{d#1}}
\newcommand{\Tr}[1]{\operatorname{Tr #1}}
\newcommand{\diag}[1]{\operatorname{diag}(#1)}
\newcommand{\tr}[1]{\operatorname{tr}#1}
\newcommand{\ad}[1]{\operatorname{ad}#1}

\begin{document}

\begin{titlepage}
\author{Martin Wolf}
\begin{center}
 {\Large\bf Soliton-Antisoliton Scattering Configurations in a Noncommutative Sigma Model in 2+1 Dimensions}
\end{center}
 
\vspace*{0.4in}

\begin{center}
  {\large Martin Wolf}\\[0.3in]
  {\em Department of Physics\\            
       Duke University, Box 90305\\                      
       Durham, North Carolina 27708-0305, USA\\[0.2in]
       
       Institut f\"ur Theoretische Physik\\
       Universit\"at Hannover\\
       30167, Hannover, Germany\\[0.2in]
       
       Institut f\"ur Theoretische Physik\\
       Technische Universit\"at Dresden\\
       01062, Dresden, Germany}
 \end{center}
\begin{center}
 Email: martin@physik.phy.tu-dresden.de
\end{center}
\vspace*{0.2in}

\begin{abstract}
\noindent
In this paper we study the noncommutative extension of a modified $U(n)$ sigma model in $2+1$ dimensions. Using the method of dressing transformations, an iterative approach for the construction of solutions from a given seed solution, we demonstrate the construction of multi-soliton and soliton-antisoliton configurations for general $n$. As illustrative examples we discuss $U(3)$ solitons and consider the head-on collision of a $U(2)$ soliton and an antisoliton explicitly, which will result in a $90^{\text{o}}$ angle scattering. Further we discuss the head-on collision of one $U(2)$ soliton with two antisolitons. This results in a $60^{\text{o}}$ angle scattering.
\end{abstract}

\end{titlepage}


\section{Introduction}

Noncommutativity in mathematics and physics has a long tradition and plays an important role. For instance, the central mathematical concept in expressing the uncertainty principle in quantum mechanics is based on noncommutativity which is reflected in any pair of conjugate variables. 

Imposing noncommutativity on spacetime is an old idea \cite{Snyder:1947qz} and offers one way to introduce nonlocality into the area of (quantum) field theory in manageable manner. The studies of noncommutative field theories were lately stimulated by the discovery that quantum field theory on a noncommutative spacetime naturally arises in a certain zero-slope limit of effective field theories of open string modes coupled to a two-form background field ($B$-field) \cite{Seiberg:1999vs}.

 Already the classical theories exhibit a rich mathematical and physical structure which is worthwhile to study. They provide us with a variety of non-perturbative solutions, such as instantons and solitons, which may be interpreted as $D$-branes in a string theory context (see reviews \cite{Nekrasov:2000ih} - \cite{Konechny:2001wz} and references therein). The analysis of solitonic solutions and their scattering properties in various field theories has made a lot of advancements (see e.g. \cite{Gopakumar:2000zd} - \cite{Furuta:2002ty} and \cite{Nekrasov:2000ih} - \cite{Konechny:2001wz}).

More than a decade ago it was shown \cite{Ooguri:1990ww}, \cite{Ooguri:1991fp} that the open $N=2$ fermionic string is identical (at tree level) to self-dual Yang-Mills theory in $2+2$ dimensions. Switching on a constant $B$-field background generates a noncommutative self-dual Yang-Mills theory on the target space. Furthermore it has been proven \cite{Lechtenfeld:2001uq}, \cite{Lechtenfeld:2000nm} that in this case the $B$-field background induces on the world volume of $n$ coincident $D2$-branes a noncommutative generalization of a modified $U(n)$ sigma model in $2+1$ dimensions, which in the commutative setup was introduced by Ward \cite{Ward:ie}.

In this paper we generalize the ideas given in \cite{Lechtenfeld:2001gf} of the construction of multi-soliton solutions of the noncommutative modified sigma model mentioned above to general $n$ as well as present the construction of soliton-antisoliton configurations using the method of dressing tranformations. 

The organization of this paper is as follows. First we briefly introduce the commutative sigma model in 2+1 dimensions, present its field equations and the energy functional. We then recall its noncommutative extension. The dressing method is outlined in section $3$ as well as its application to the construction of non-Abelian multi-soliton configurations. In section $4$ we derive the non-Abelian soliton-antisoliton solutions.


\section{Commutative vs. noncommutative modified sigma model}

{\bf Commutative model.} It is well known, that nonlinear sigma models in $2+1$ dimensional Minkowski space may be Lorentz-invariant or integrable, however, not both. In this paper we shall choose the property of integrability and discuss a noncommutative extension of a modified $U(n)$ sigma model introduced by Ward \cite{Ward:ie}\footnote{The original model is a $SU(n)$ model. However, in the noncommutative extension we need to go from $SU(n)$ to $U(n)$ \cite{Terashima:2000xq}, \cite{Matsubara:2000gr}.}. His model describes the dynamics of an $U(n)$-valued field $\Phi:\mathbb{R}^{1,2}\rightarrow U(n)$. Here $\mathbb{R}^{1,2}$ is the physical coordinate space $(t,x,y)$ with the usual metric $(\eta_{\mu\nu})=\diag{-1,1,1}$.

The field equations are given by
\begin{equation}\label{fe-cv}
 \eta^{\mu\nu}\partial_{\mu}(\Phi^{-1}\partial_{\nu}\Phi)
 +V_{\alpha}\epsilon^{\alpha\mu\nu}\partial_{\mu}(\Phi^{-1}\partial_{\nu}\Phi)=0,
\end{equation}
with $(V_{\alpha})=(0,1,0)$, $\epsilon^{\alpha\mu\nu}$ the Levi-Civita symbol and $\partial_{\mu}$ denotes the partial derivative with respect to $x^{\mu}$. The second term breaks the Lorentz group of $SO(1,2)$ explicitly to $GL(1,\mathbb{R})$ generated by boosts in $y$-direction. However, (\ref{fe-cv}) is integrable \cite{Ward:ie} and admits scattering multi-soliton configurations \cite{Ward:1995zy}. 

This model exhibits a conserved energy functional
\begin{equation}\label{e-cv}
 E=\int\limits_{\mathbb{R}^2}\D{ ^2x}\,\mathcal{E}=\frac{1}{2}\int\limits_{\mathbb{R}^2}\D{ ^2x}\,\tr{}\left\{\partial_t\Phi^{-1}\partial_t\Phi+\partial_x\Phi^{-1}\partial_x\Phi+\partial_y\Phi^{-1}\partial_y\Phi\right\},
\end{equation}
where `$\tr{}$' implies the trace over the $U(n)$ group space. Note, that the energy density $\mathcal{E}$ defined in (\ref{e-cv}) is a positive semidefinite functional of the field $\Phi$. In the noncommutative extension this will no longer hold.\\

\noindent {\bf Noncommutative model.} The usual procedure of a noncommutative extension is the deformation of the pointwise product between functions via the star (or Moyal) product defined by
\begin{equation}\label{sp}
 (f\star g)(x):=f(x)\,\exp\left\{\frac{i}{2}\overleftarrow{\partial_{\mu}}\theta^{\mu\nu}\overrightarrow{\partial_{\nu}}\right\}\,g(x)\qquad\text{for}\qquad\mu,\nu=1,2,3,
\end{equation}
with $x\in\mathbb{R}^{1,2}$, $f,g\in C^{\infty}(\mathbb{R}^{1,2})$ and $\theta^{\mu\nu}$ is chosen to be a constant antisymmetric matrix. Equation (\ref{sp}) implies that
\begin{equation}
 [x^{\mu},x^{\nu}]_{\star}:=x^{\mu}\star x^{\nu}-x^{\nu}\star x^{\mu}=i\theta^{\mu\nu}\qquad\text{for}\qquad\mu,\nu=1,2,3.
\end{equation}
In the following we assume a spatial noncommutativity only, such that
\begin{equation}
 \theta^{23}=-\theta^{32}=:\theta>0,
\end{equation}
with all other components of $\theta^{\mu\nu}$ equal to zero and hence,
\begin{equation}
 [x,y]_{\star}=i\theta.
\end{equation}

It is convenient to perform a change of coordinates, namely
\begin{equation}
 u:=\frac{1}{2}(t+y),\qquad v:=\frac{1}{2}(t-y),\qquad\partial_u=\partial_t+\partial_y,\qquad\partial_v=\partial_t-\partial_y.
\end{equation}
Then the noncommutative field equations simply read
\begin{equation}\label{fe-ncv}
 \partial_x(\Phi^{-1}\star\partial_x\Phi)-\partial_v(\Phi^{-1}\star\partial_u\Phi)=0
\end{equation}
and the conserved energy functional
\begin{equation}\label{e-ncv}
 E=\int\limits_{\mathbb{R}^2}\D{ ^2x}\,\mathcal{E}_{\star}=\frac{1}{2}\int\limits_{\mathbb{R}^2}\D{ ^2x}\,\tr{}\left\{\partial_t\Phi^{-1}\star\partial_t\Phi+\partial_x\Phi^{-1}\star\partial_x\Phi+\partial_y\Phi^{-1}\star\partial_y\Phi\right\}.
\end{equation}
Note, that the energy density $\mathcal{E}_{\star}$ defined above need not necessarily to be positive semidefinite, since even for real-valued functions $f$, $f\star f$ is real but not always positive; the total energy, however, remains positive semidefinite.

The non-locality of the star product causes often difficulties in explicit calculations. It is therefore helpful to pass over to the operator formalism, i.e.
\begin{equation}
 [t,\hat{x}]=[t,\hat{y}]=0,\qquad[\hat{x},\hat{y}]=i\theta\quad\text{and}\quad[\hat{z},\hat{\bar{z}}]=2\theta,
\end{equation}
with $\hat{z}=\hat{x}+i\hat{y}$ and $\hat{\bar{z}}=\hat{x}-i\hat{y}$. Since the last of these equations has the form of the Heisenberg algebra we introduce creation and annihilation operators
\begin{equation}
 a=\frac{1}{\sqrt{2\theta}}\hat{z}\quad\text{and}\quad a^\dagger=\frac{1}{\sqrt{2\theta}}\hat{\bar{z}}
   \quad\text{with}\quad[a,a^\dagger]=1,
\end{equation}
acting on the harmonic oscillator Fock space $\mathcal{H}=\bigoplus_n\mathbb{C}\,|n\rangle$ with $\langle m|n \rangle=\delta_{mn}$ such that
\begin{equation}
 a^\dagger a|n\rangle=:N|n\rangle=n|n\rangle,\quad a|n\rangle=\sqrt{n}|n-1\rangle\quad\text{and}\quad 
 a^\dagger |n\rangle=\sqrt{n+1}|n+1\rangle.
\end{equation}

\noindent {\bf Weyl transform.} The linear map $\mathcal{W}$ which transforms ordinary functions $f(t,z,\bar{z})$ into operator-valued functions $\hat{f}(t):=\mathcal{O}_f(t)$ acting on $\mathcal{H}$ is called the Weyl transform and defined by
\begin{eqnarray}
 f(t,z,\bar{z}) &\mapsto& \mathcal{O}_f(t)=\mathcal{W}\{f(t,z,\bar{z})\}\notag\\
                &=:&      -\int\limits_{\mathbb{C}\times\mathbb{C}}\frac{\D{p}\D{\bar{p}}}{(2\pi)^2}
                          \D{z}\D{\bar{z}}\,f(t,z,\bar{z})\,e^{-i[\bar{p}(\sqrt{2\theta}a-z)
                          +p(\sqrt{2\theta}a^\dagger-\bar{z})]}.\label{weyl}
\end{eqnarray}
The inverse transform is given by
\begin{eqnarray}
 \mathcal{O}_f(t) &\mapsto& f(t,z,\bar{z})=\mathcal{W}^{-1}\{\mathcal{O}_f(t)\}\notag\\
                  &=&       2\pi\theta\,\int\limits_{\mathbb{C}}\frac{2i\D{p}\D{\bar{p}}}{(2\pi)^2}\,\Tr{}
                            \left\{\mathcal{O}_f(t)\,e^{i[\bar{p}(\sqrt{2\theta}a-z)
                            +p(\sqrt{2\theta}a^\dagger-\bar{z})]}\right\},\label{inv-weyl}
\end{eqnarray}
where `$\Tr{}$' denotes the trace over the Fock space $\mathcal{H}$. Further we have some useful properties
\begin{equation}
 \mathcal{O}_{f\star g}=\mathcal{O}_f\mathcal{O}_g\qquad\text{and}\qquad
 \int\limits_{\mathbb{R}^2}\D{ ^2x}\,f=2\pi\theta\Tr{}\mathcal{O}_f
\end{equation}
as well as
\begin{equation}
 \hat{\partial}_x\hat{f}:=\mathcal{O}_{\partial_xf}=\frac{i}{\theta}\ad{(\hat{y})}(\mathcal{O}_f)
                            \qquad\text{and}\qquad
 \hat{\partial}_y\hat{f}:=\mathcal{O}_{\partial_yf}=-\frac{i}{\theta}\ad{(\hat{x})}(\mathcal{O}_f)
\end{equation}
such that,
\begin{equation}
 \hat{\partial}_z\hat{f}:=\mathcal{O}_{\partial_zf}=\frac{-1}{\sqrt{2\theta}}\ad{(a^\dagger)}(\mathcal{O}_f)
                            \qquad\text{and}\qquad
 \hat{\partial}_{\bar{z}}\hat{f}:=\mathcal{O}_{\partial_{\bar{z}}f}=\frac{1}{\sqrt{2\theta}}
                                  \ad{(a)}(\mathcal{O}_f).
\end{equation}
Using creation and annihilation operators the energy density defined in (\ref{e-ncv}) is given by
\begin{equation}\label{e-ncv-ca}
 \hat{\mathcal{E}}=\frac{1}{2}\tr{}\{\partial_t\hat{\Phi}^\dagger\partial_t\hat{\Phi}\}+\frac{1}{2\theta}\tr{}\{
  [a,\hat{\Phi}^\dagger][a,\hat{\Phi}^\dagger]^\dagger+[a^\dagger,\hat{\Phi}^\dagger][a^\dagger,\hat{\Phi}^\dagger]^\dagger\},
\end{equation}
where `$\tr{}$' denotes the trace over the $U(n)$ group space. The total energy is then obtained by computing the trace over the Fock space $\mathcal{H}$, that is
\begin{equation}\label{e-ncv-tot}
 E=2\pi\theta\Tr{}\hat{\mathcal{E}}.
\end{equation}
In order to avoid an abuse of notation we will from now on omit the hats over the operators except when confusion may arise.


\section{Dressing approach and soliton configurations}

In this section we are going to generalize the soliton solutions presented in \cite{Lechtenfeld:2001gf} to general $n>1$ using the so-called dressing method, which is a recursive procedure of generating solutions from a given seed solution. Originally, this method had been developed for commutative integrable systems (see e.g. \cite{zak}, \cite{Forgacs:1983gr} and \cite{Ward:ie}). However, its transition to the noncommutative situation is readily accomplished \cite{Lechtenfeld:2001uq}. Before we focus on the modified sigma model, however, let us briefly recall this method.\\

\noindent {\bf Linear system.} Starting point are the two linear partial differential equations
\begin{equation}\label{l-pdes}
 (\zeta\partial_x-\partial_u)\psi=A\psi\qquad\text{and}\qquad (\zeta\partial_v-\partial_x)\psi=B\psi,
\end{equation}
where $\zeta$ is called the spectral parameter with $\zeta\in\mathbb{C}\cup\{\infty\}\cong\mathbb{CP}^1$. The matrix-valued function $\psi$ depends on $(t,x,y,\zeta)$ (or equivalently on $(x,u,v,\zeta)$), i.e. $\psi:\mathbb{R}^{1,2}\times\mathbb{CP}^1\rightarrow\text{Mat}_n(\mathcal{H})$. Here $\text{Mat}_n(\mathcal{H})$ is the set of all $n\times n$ matrices whose elements are operators acting on $\mathcal{H}$. The matrices $A$ and $B$ are elements of $\text{Mat}_n(\mathcal{H})$ as well, but do not depend on the spectral parameter $\zeta$. According to \cite{Ward:ie} we impose the following reality condition on $\psi$:
\begin{equation}\label{rc}
 \psi(t,x,y,\zeta)[\psi(t,x,y,\bar{\zeta})]^\dagger=1.
\end{equation}
The integrability conditions for the system (\ref{l-pdes}) are given by
\begin{eqnarray}
 \partial_xB-\partial_vA       &=& 0,\\
 \partial_xA-\partial_uB-[A,B] &=& 0.
\end{eqnarray}
Obviously, $A=\Phi^{-1}\partial_u\Phi$ and $B=\Phi^{-1}\partial_x\Phi$ solve the second equation and transform the first one into 
\begin{equation}
 \partial_x(\Phi^{-1}\partial_x\Phi)-\partial_v(\Phi^{-1}\partial_u\Phi)=0,
\end{equation}
which is exactly the operator version of (\ref{fe-ncv}). The standard asymptotic conditions \cite{Ivanova:2000zt} read
\begin{eqnarray}
 \psi(t,x,y,\zeta\to\infty) &=& 1+O(\zeta^{-1}),\\
 \psi(t,x,y,\zeta\to 0)     &=& \Phi^{-1}(t,x,y)+O(\zeta).
\end{eqnarray}
Note, that using (\ref{rc}) the equations (\ref{l-pdes}) can be rewritten  as
\begin{eqnarray}
 -\psi(t,x,y,\zeta)(\zeta\partial_x-\partial_u)[\psi(t,x,y,\bar{\zeta})]^\dagger &=& A(t,x,y),\\
 -\psi(t,x,y,\zeta)(\zeta\partial_v-\partial_x)[\psi(t,x,y,\bar{\zeta})]^\dagger &=& B(t,x,y),
\end{eqnarray}
i.e. knowing the auxiliary field $\psi$ we can construct $\Phi$, $A$ and $B$, respectively.\\

\noindent {\bf Ansatz and dressing method.} In \cite{Lechtenfeld:2001aw} and \cite{Lechtenfeld:2001gf} solutions of (\ref{l-pdes}) of the form
\begin{equation}\label{ansatz-dressing}
 \tilde{\psi}(t,x,y,\zeta)=\chi(t,x,y,\zeta)\psi(t,x,y,\zeta)\quad\text{with}\quad\chi=1+\sum_{\alpha=1}^{s}
                           \sum_{k=1}^{m}\frac{R_{\alpha k}}{(\zeta-\mu_k)^\alpha},
\end{equation}
have been considered with $\psi$ a given seed solution of (\ref{l-pdes}), $\mu_k\in\mathbb{C}$ and the operator-valued matrices $R_{\alpha k}:\mathbb{R}^{1,2}\rightarrow\text{Mat}_n(\mathcal{H})$ are independent of $\zeta$. The operator-valued matrix $\chi$ is called the dressing factor of the dressing transformation. In the following we restrict ourselves to first order poles in the dressing factor $\chi$, i.e. $s=1$ in (\ref{ansatz-dressing}).

First, we choose $\psi=1$. Then we find
\begin{equation}
 \psi_0=1+\sum_{k=1}^{m}\frac{R_k}{\zeta-\mu_k},
\end{equation}
with $\psi_0:=\tilde{\psi}$. Recall, that the matrices $R_k$ are given via $n\times r$ matrices $T_k:\mathbb{R}^{1,2}\rightarrow\text{Mat}_{n\times r}(\mathcal{H})$ \cite{Lechtenfeld:2001aw}, \cite{Lechtenfeld:2001gf}. Further, it was shown that the simplest case occurs when $\psi_0$ has only one pole at $\zeta=-i$,
\begin{equation}\label{static}
 \psi_0=1+\frac{R}{\zeta+i}=:1-\frac{2i}{\zeta+i}P\qquad\text{and hence,}\qquad\Phi=1-2P.
\end{equation}
In that case all configurations are static and parametrized by a Hermitian projector $P=T(T^\dagger T)^{-1}T^\dagger$ which satifies
\begin{equation}\label{s-sol}
 (1-P)\partial_{\bar{z}}P=0\qquad\Rightarrow\qquad (1-P)aP=0.
\end{equation}
Then $T$ obeys the equation
\begin{equation}
 (1-P)aT=0,
\end{equation}
what means that $aT$ lies in the kernel of $1-P$.

Now we choose the static configuration (\ref{static}) to be the seed solution and consider a dressing transformation with $\chi$ being of the same form, i.e.
\begin{equation}
 \psi_0\mapsto \tilde{\psi}=\chi\psi_0=
                       \biggl(1-\frac{2i}{\zeta+i}\tilde{P}\biggr)\biggl(1-\frac{2i}{\zeta+i}P\biggr),
\end{equation}
where $\tilde{P}$ is some matrix. This leads to a configuration of the form
\begin{equation}
 \Phi^{-1}=\tilde{\psi}(\zeta\to 0)=(1-2\tilde{P})(1-2P).
\end{equation}
Note, that $\tilde{\psi}$ contains a second-order pole at $\zeta=-i$. The reality condition (\ref{rc}) implies that 
\begin{equation}
 \tilde{P}^\dagger=\tilde{P}\qquad\text{and}\qquad\tilde{P}^2=\tilde{P},
\end{equation}
what means that $\tilde{P}$ can also be considered to be a Hermitian projector $\tilde{P}=\tilde{T}(\tilde{T}^\dagger\tilde{T})^{-1}\tilde{T}^\dagger$ with some $n\times\tilde{r}$ matrix $\tilde{T}$.\\

\noindent {\bf Linear equations}. Demanding that $\tilde{\psi}$ is again a solution of (\ref{l-pdes}) it was shown \cite{Lechtenfeld:2001gf} that $\tilde{T}$ has to satisfy 
\begin{equation}\label{sol-sol}
 a\tilde{T}+[a,P]\tilde{T}=\tilde{T}Z_1\qquad\text{and}\qquad\partial_t\tilde{T}+i\gamma[a^\dagger,P]\tilde{T}=\tilde{T}Z_2,
\end{equation}
where $\gamma:=\sqrt{2/\theta}$ and $Z_{1,2}(t,a,a^\dagger)$ are some operator-valued functions.\\

\noindent {\bf Non-Abelian multi-soliton configurations.} In \cite{Lechtenfeld:2001gf} the solutions were restricted to the $U(2)$ group. The seed configuration was taken to be the simplest non-trivial solution of (\ref{s-sol}), namely
\begin{equation}\label{ansatz-u(2)}
 P = T\frac{1}{T^\dagger T}T^\dagger
   = \begin{pmatrix} 
      \frac{1}{1+\bar{z}z}   &  \frac{1}{1+\bar{z}z}\bar{z}\\ 
      z\frac{1}{1+\bar{z}z}  & z\frac{1}{1+\bar{z}z}\bar{z}
     \end{pmatrix}
 \quad\text{with}\quad
 T = \begin{pmatrix} 1 \\ z \end{pmatrix}
 \quad\text{and}\quad \bar{z}\equiv z^\dagger.
\end{equation}
The total energy of the configuration (\ref{ansatz-u(2)}) is readily computed to be $E=8\pi$ using (\ref{e-ncv-ca}) and (\ref{e-ncv-tot}). Let us now generalize this ansatz to $U(n)$ with $n>1$. For notational simplicity we define
\begin{equation}
 K :=\frac{1}{1+\sum_{i=1}^{n-1}\bar{z}^{m_i}z^{m_i}}\quad\text{with}\quad m_i\in\mathbb{N}\cup\{0\}\quad
     \text{for}\quad i=1,\ldots, n-1.
\end{equation}
Then $P$ takes the following form
\begin{equation}\label{ansatz-u(n)}
 P = T\frac{1}{T^\dagger T}T^\dagger
     = \begin{pmatrix}
        K             & K\bar{z}^{m_1}             &  \cdots  & K\bar{z}^{m_{n-1}}\\
        z^{m_1}K      & z^{m_1}K\bar{z}^{m_1}      &  \cdots  & z^{m_1}K\bar{z}^{m_{n-1}}\\
        \vdots        & \vdots                     &  \ddots  & \vdots\\
        z^{m_{n-1}}K  & z^{m_{n-1}}K\bar{z}^{m_1}  &  \cdots  &  z^{m_{n-1}}K\bar{z}^{m_{n-1}}
       \end{pmatrix},
\end{equation}
with $T^\dagger=(1,\bar{z}^{m_1},\ldots,\bar{z}^{m_{n-1}})$. This ansatz describes a field configuration consisting of $q:=\max_{i}\{m_i\}$ static lumps. Again, with (\ref{e-ncv-ca}) and (\ref{e-ncv-tot}) the total energy is given by
\begin{equation}
 E=8\pi q=8\pi\max_{i}\{m_i\}.
\end{equation}
We build the dressing factor $\chi$ with a matrix $\tilde{T}^\dagger=(\bar{u}_1,\ldots,\bar{u}_n)$, where $u_i(t,z,\bar{z})$ are operator-valued functions to be determined. Choosing $Z_1=a$ and the ansatz for $\tilde{T}$ the first of the equations (\ref{sol-sol})  turns into
\begin{eqnarray}
 [z,u_1]       &=& [K,z]\biggl(u_1+\sum_{i=1}^{n-1}\bar{z}^{m_i}u_{i+1}\biggr)-
                   2\theta K\sum_{i=1}^{n-1}m_i\bar{z}^{m_i-1}u_{i+1},\\
 {[z,u_{i+1}]} &=& z^{m_i}[K,z]\biggl(u_1+\sum_{i=1}^{n-1}\bar{z}^{m_i}u_{i+1}\biggr)-
                   z^{m_i}2\theta K\sum_{i=1}^{n-1}m_i\bar{z}^{m_i-1}u_{i+1},
\end{eqnarray}
for $i=1,\ldots,n-1$. These equations immediately tell us that
\begin{equation}\label{sol-eq1}
 [z,z^{m_i}u_1-u_{i+1}]=0\qquad\Rightarrow\qquad u_{i+1}=z^{m_i}u_1-f_{i+1}(t,z),
\end{equation}
whereby $f_{i+1}(t,z)$ is arbitrary, but independent of $\bar{z}$. Imposing $Z_2=0$ the second equation of (\ref{sol-sol}) reduces to
\begin{equation}\label{sol-eq2}
 i\theta\partial_tu_1 = [\bar{z},K]\biggl(u_1+\sum_{i=1}^{n-1}\bar{z}^{m_i}u_{i+1}\biggr).
\end{equation}
Plugging relation (\ref{sol-eq1}) into (\ref{sol-eq2}) and defining
\begin{equation}
 u_1=1+K\sum_{i=1}^{n-1}\bar{z}^{m_i}f_{i+1}(t,z)
\end{equation}
yields
\begin{equation}
 i\theta\partial_tu_1=[\bar{z},K]K^{-1}.
\end{equation}
It is not difficult to see that
\begin{equation}
 f_{i+1}(t,z)=-2i(m_iz^{m_i-1}t+h_{i+1}(z))
\end{equation}
solves our equations consistently, where the $h_{i+1}$'s (for $i=1,\ldots,n-1$) are arbitrary holomorphic operator-valued functions which depend only on $z$. Hence,
\begin{equation}
 \tilde{T}=\begin{pmatrix} 1\\ z^{m_1}\\ \vdots \\ z^{m_{n-1}} \end{pmatrix} +
           \begin{pmatrix} 1\\ z^{m_1}\\ \vdots \\ z^{m_{n-1}} \end{pmatrix}\,
           \frac{1}{1+\sum_{i=1}^{n-1}\bar{z}^{m_i}z^{m_i}}\,\sum_{i=1}^{n-1}\bar{z}^if_{i+1}-
           \begin{pmatrix} 0\\ f_2    \\ \vdots \\ f_n         \end{pmatrix}.
\end{equation}
Note, that for locations given by $m_iz^{m_i-1}t+h_{i+1}(z)=0$ for all $i=1,\ldots, n-1$ our solution degenerates, since $\tilde{T}=T$ and therefore $\Phi=(1-2P)^2=1$.\\

\noindent {\bf Examples.} Obviously, these solutions do agree for $n=2$ with those given in \cite{Lechtenfeld:2001gf}. Now let us restrict to $n=3$. For this case $\tilde{T}$ turns out to be
\begin{equation}
 \tilde{T}=\begin{pmatrix} 1\\ z^{m}\\ z^{n} \end{pmatrix} +
           \begin{pmatrix} 1\\ z^{m}\\ z^{n} \end{pmatrix}\,
            \frac{1}{1+\bar{z}^{m}z^{m}+\bar{z}^{n}z^{n}}\,(\bar{z}^{m}f+\bar{z}^{n}g)-
           \begin{pmatrix} 0\\ f\\ g \end{pmatrix},
\end{equation}
with $m:=m_1$ and $n:=m_2$, $f:=f_2$ and $g:=f_3$ and hence,
\begin{equation}
 f=-2i(mz^{m-1}t+h_2(z))\qquad\text{and}\qquad g=-2i(nz^{n-1}t+h_3(z)).
\end{equation}
Performing the inverse Weyl transform (\ref{inv-weyl}) yields
\begin{equation}\label{solsol-u(3)}
 \tilde{T}\mapsto\tilde{T}_{\star} 
   = \begin{pmatrix} 1\\ z^{m}\\ z^{n} \end{pmatrix} +
     \begin{pmatrix} 1\\ z^{m}\\ z^{n} \end{pmatrix}
        \star\,\frac{1}{1+\bar{z}^{m}\star z^{m}+\bar{z}^{n}\star z^{n}}\,
        \star(\bar{z}^{m}\star f_{\star}+\bar{z}^{n}\star g_{\star})-
     \begin{pmatrix} 0\\ f_{\star}\\ g_{\star} \end{pmatrix},
\end{equation}
with the commutative coordinates $t,z,\bar{z}$. While $P_{\star}$ is given by $T_{\star}$, we can calculate the projector $\tilde{P}_{\star}$ using (\ref{solsol-u(3)}), therefore the field $\Phi_{\star}=(1-2P_{\star})\star(1-2\tilde{P}_{\star})$ and hence, the energy density $\mathcal{E}_{\star}$. Obviously, the solution (\ref{solsol-u(3)}) coincides in the limit $\theta\to 0$ with the one given by Ioannidou and Zakrzewski in \cite{Ioannidou:1998jh}.

\begin{figure}[ht]
 \begin{minipage}[h]{2.5in}
  \begin{center}
   \includegraphics[height=1.5in,width=2.5in]{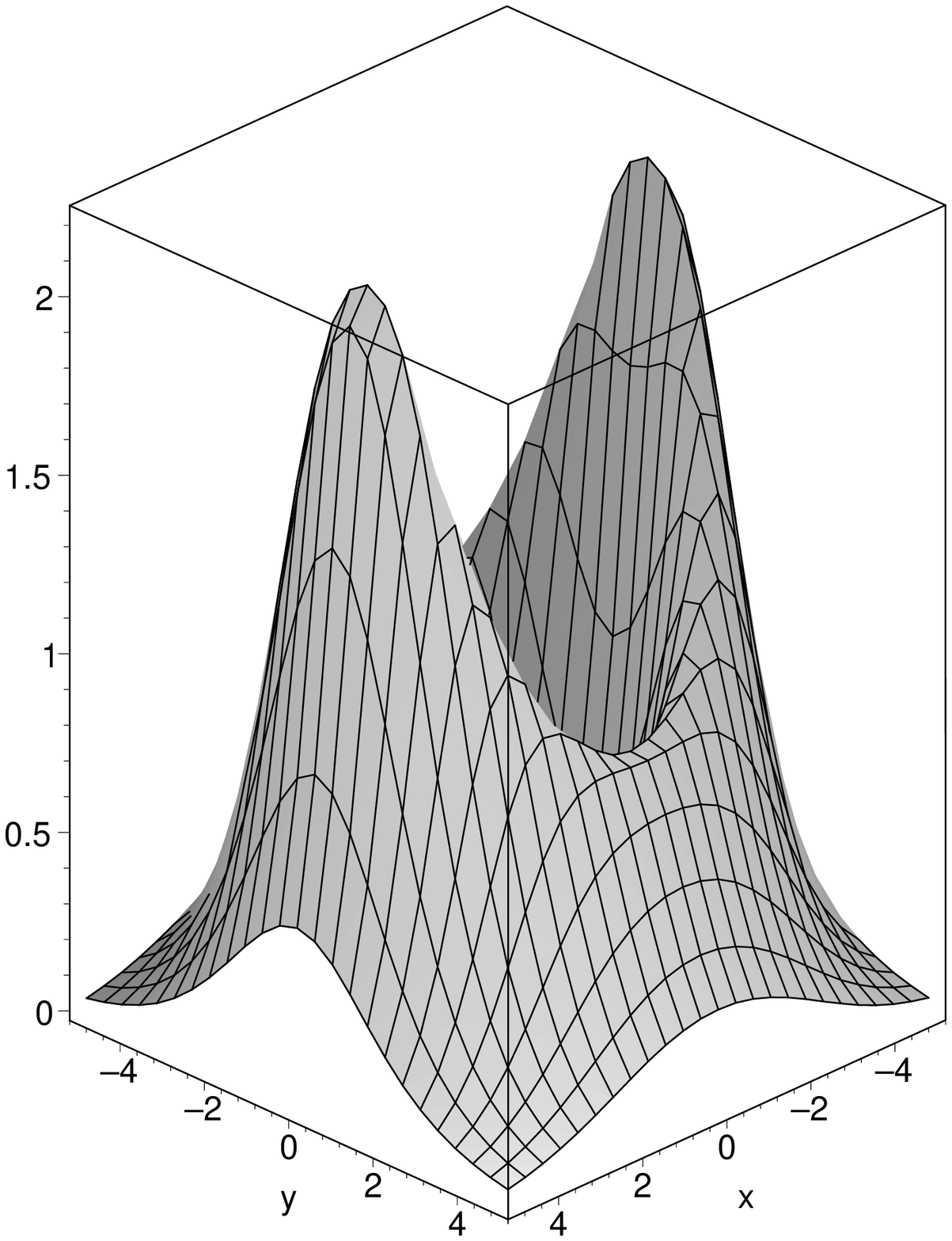}
   \caption{\small Energy density for three $U(3)$ solitons at time $t=-40$ for $\theta\ll 1$.}
   \label{ringsol1}
  \end{center}
 \end{minipage}
 \hfill
 \begin{minipage}[h]{2.5in}
  \begin{center}
   \includegraphics[height=1.5in, width=2.5in]{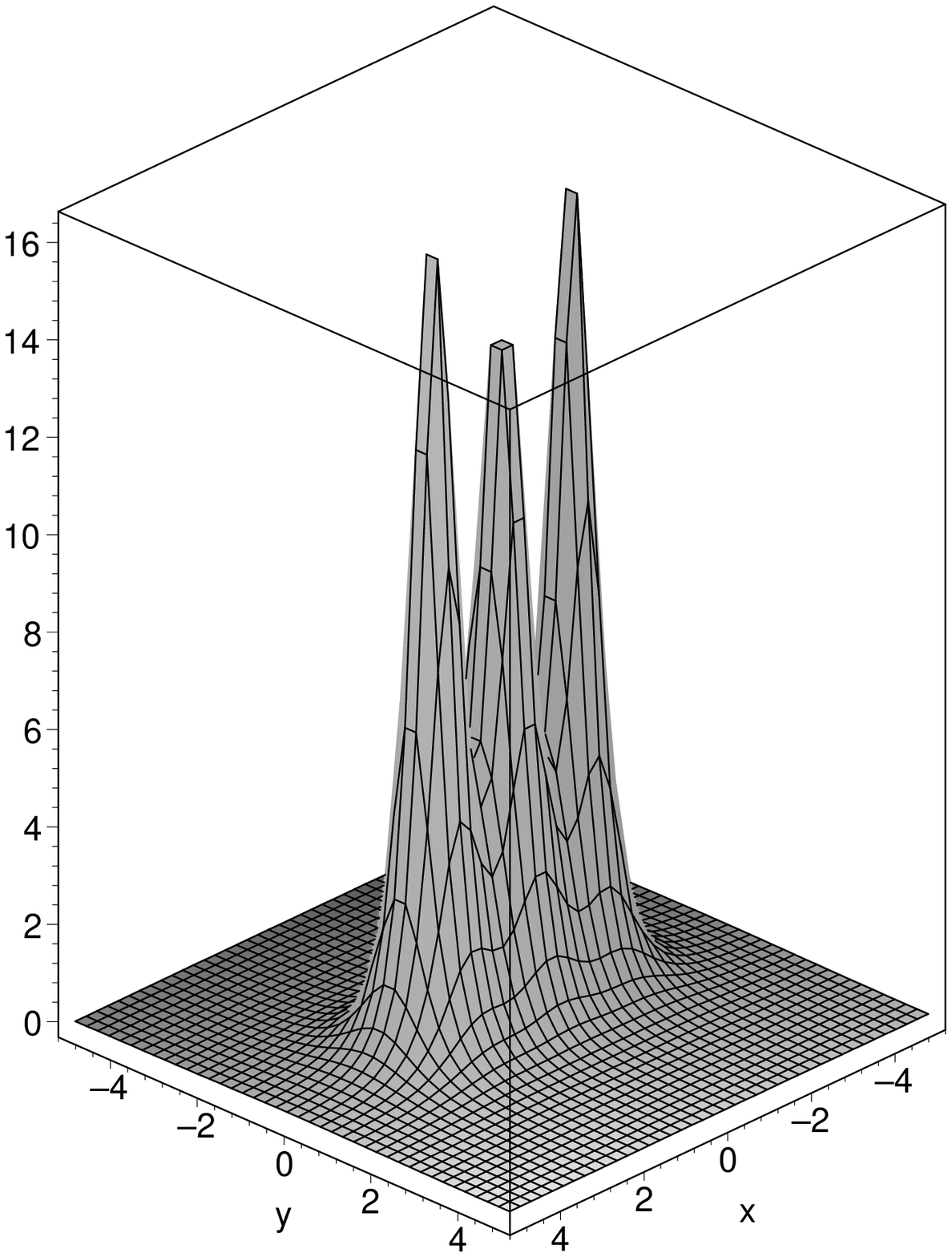}
   \caption{\small Energy density for three $U(3)$ solitons at time $t=-1$ for $\theta\ll 1$.}
   \label{ringsol2}
  \end{center}
 \end{minipage}
\end{figure}
\begin{figure}[ht]
 \begin{minipage}[h]{2.5in}
  \begin{center}
   \includegraphics[height=1.5in,width=2.5in]{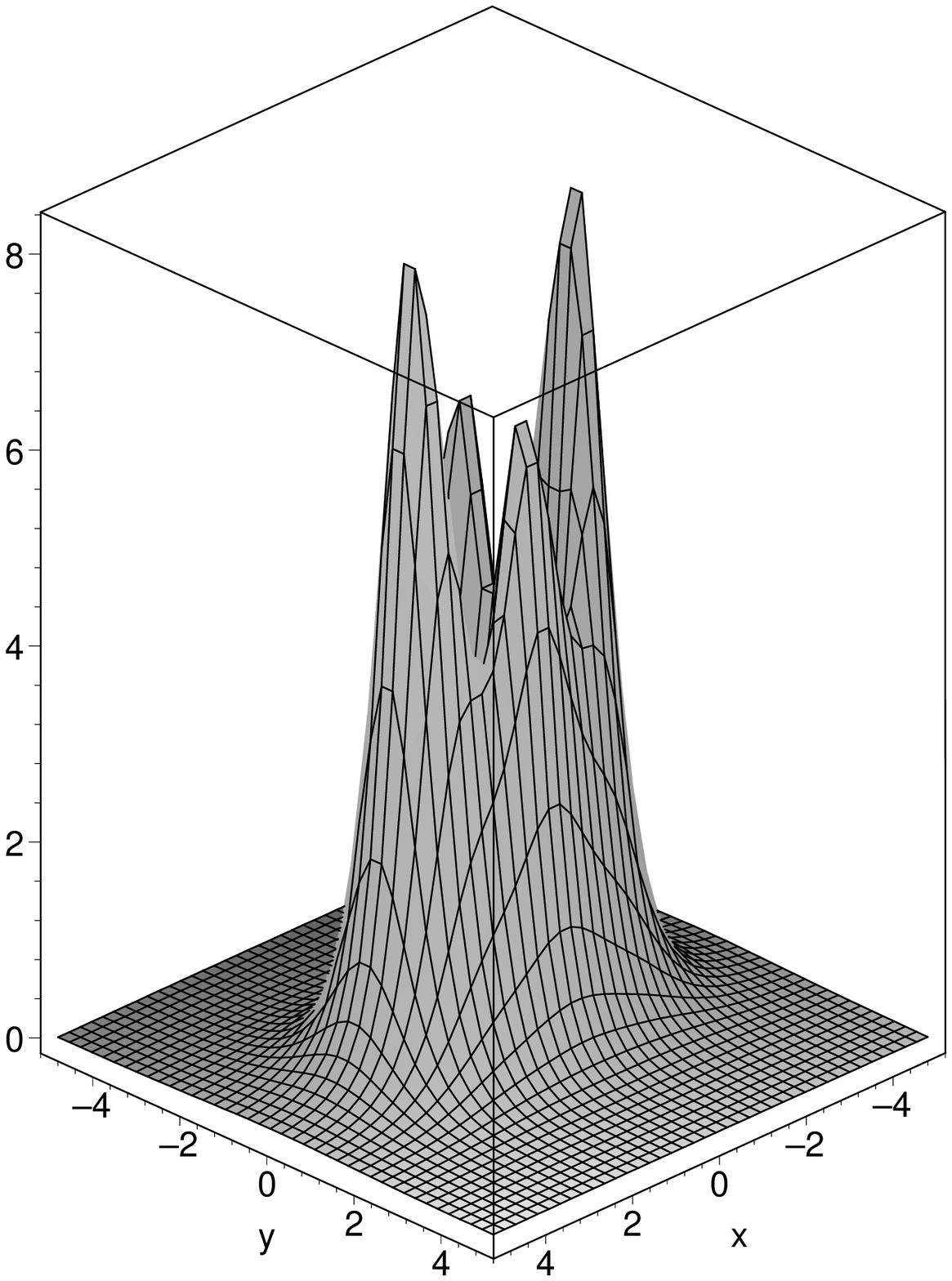}
   \caption{\small Energy density for three $U(3)$ solitons at time $t=1$ for $\theta\ll 1$.}
   \label{ringsol3}
  \end{center}
 \end{minipage}
 \hfill
 \begin{minipage}[h]{2.5in}
  \begin{center}
   \includegraphics[height=1.5in,width=2.5in]{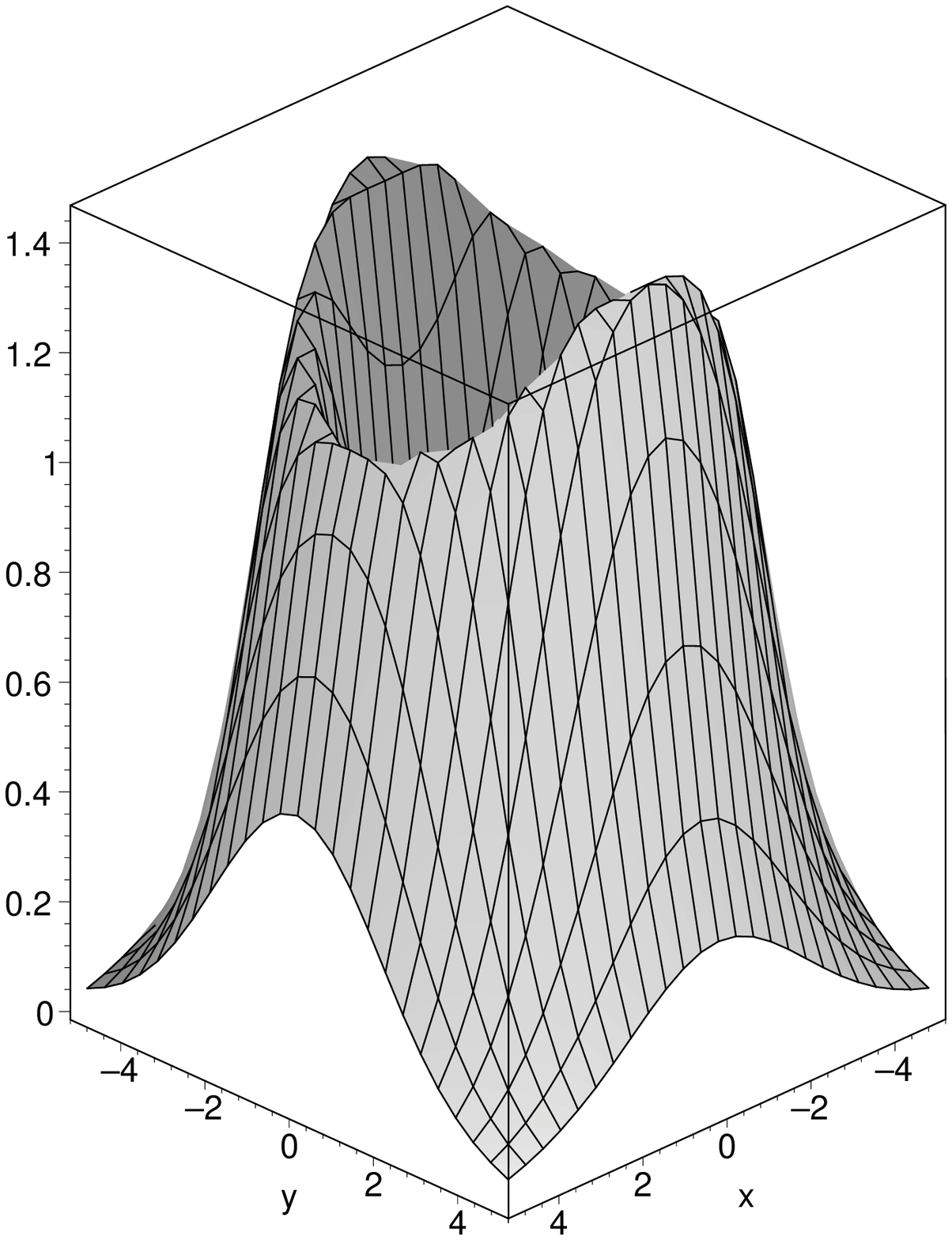}
   \caption{\small Energy density for three $U(3)$ solitons at time $t=40$ for $\theta\ll 1$.}
   \label{ringsol4}
  \end{center}
 \end{minipage}
\end{figure}

Choosing $m=n=1$, $h_2(z)=z^2$, $h_3(z)=3$ and therefore $f=-2i(t+z^2)$ and $g=-2i(t+3)$, the solution (\ref{solsol-u(3)}) represents a three soliton configuration, which for large negative and positive times has a ring-like structure. Figures \ref{ringsol1} - \ref{ringsol4} show the energy density for various times in the commutative limit.

\begin{figure}[ht]
 \begin{minipage}[h]{2.5in}
  \begin{center}
   \includegraphics[height=1.5in,width=2.5in]{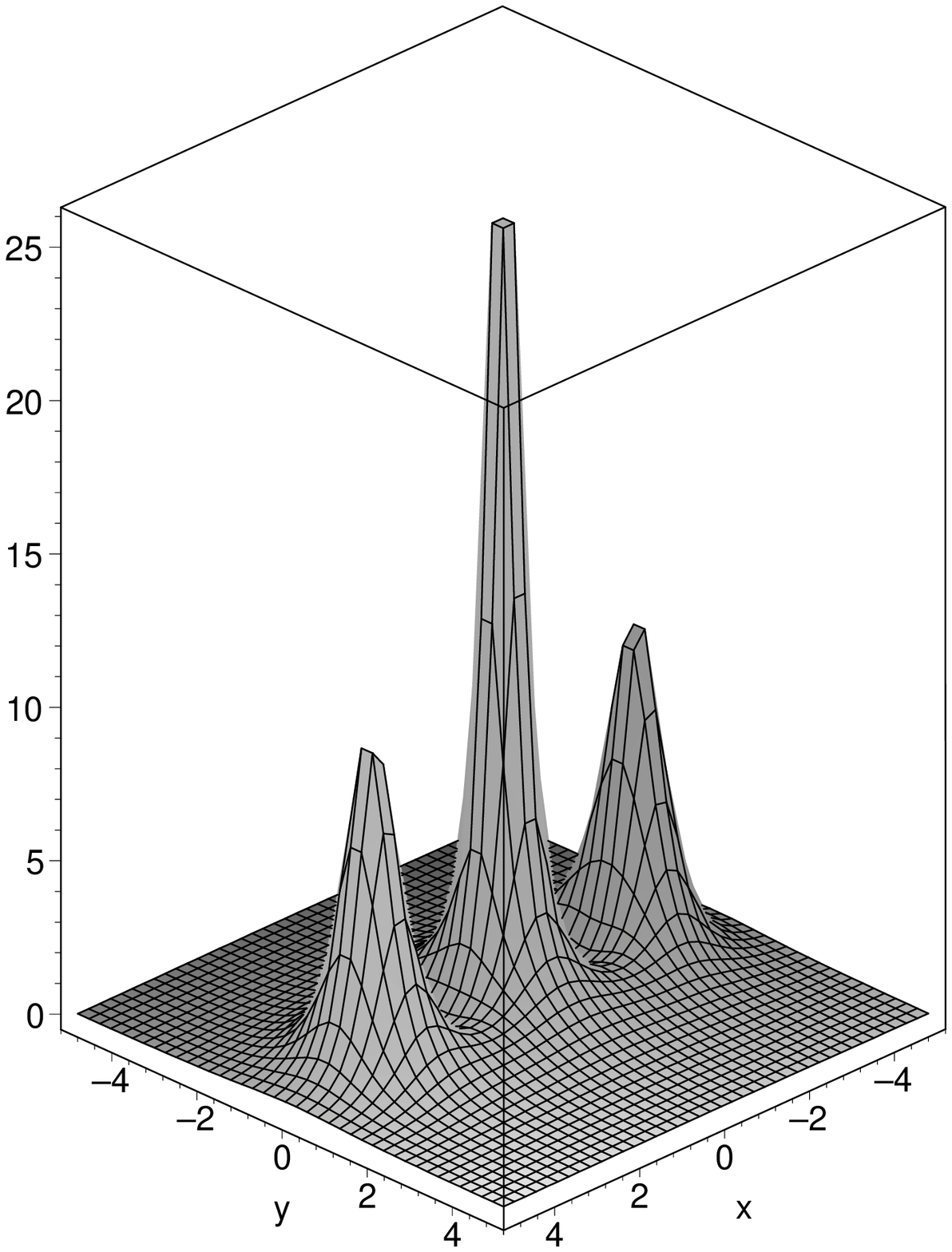}
   \caption{\small Energy density for four $U(3)$ solitons, showing right angle scattering at time $t=-10$ for 
                   $\theta\ll 1$.}
   \label{scatsol1}
  \end{center}
 \end{minipage}
 \hfill
 \begin{minipage}[h]{2.5in}
  \begin{center}
   \includegraphics[height=1.5in, width=2.5in]{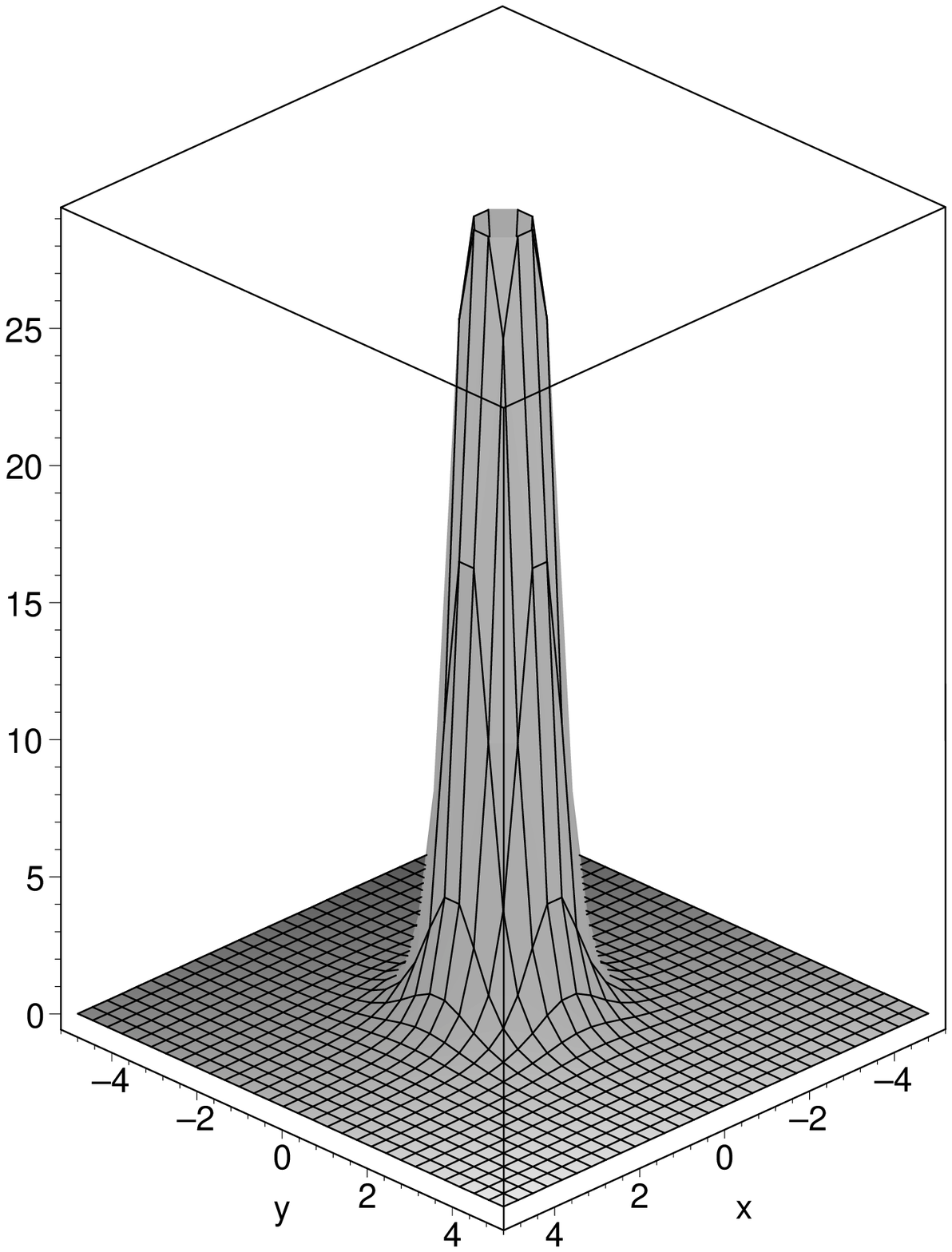}
   \caption{\small Energy density for four $U(3)$ solitons, showing right angle scattering at time $t=0$ for 
                   $\theta\ll 1$.}
   \label{scatsol3}
  \end{center}
 \end{minipage}
\end{figure}

\begin{figure}[ht]
 \begin{center}
   \includegraphics[height=1.5in,width=2.5in]{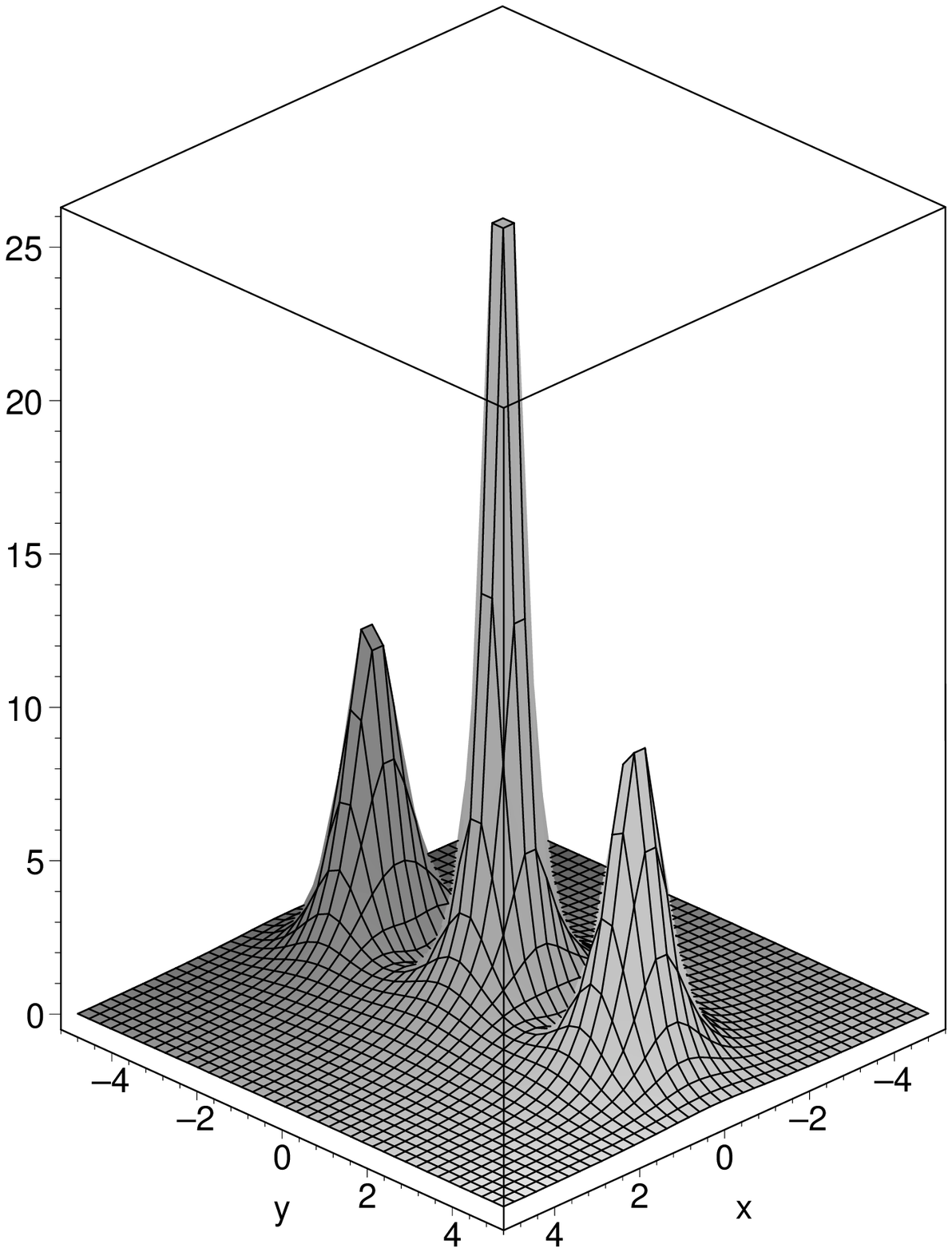}
   \caption{\small Energy density for four $U(3)$ solitons, showing right angle scattering at time $t=10$ for 
                   $\theta\ll 1$.}
   \label{scatsol5}
  \end{center}
\end{figure}

In order to obtain scattering solutions we consider $m=2$, $n=1$, $h_2(z)=z^3$ and $h_3(z)=0$. Therefore we have $f=-2i(2zt+z^3)$ and $g=-2it$. Figures \ref{scatsol1} - \ref{scatsol5} visualize the scattering configuration for small $\theta$. It consists of two static solitons sitting at the origin accompanied by two moving solitons. The latter accelerate along the $x$-axis towards the origin, perform a $90^{\text{o}}$ angle scattering and decelerate along the $y$-axis towards infinity.

We refrain from writing down the lengthy explicit expressions for the energy density (\ref{e-ncv}) for the cases discussed above.

Note, the limit $r\to\infty$, with $r^2=z\bar{z}$, is equivalent to the limit $\theta\to 0$, since similarly noncommuativity goes to zero for any fixed $\theta <\infty$. Hence, Ioannidou's and Zakrzewski's \cite{Ioannidou:1998jh} asymptotic analysis can be applied without alteration.

\section{Soliton-antisoliton configurations}

{\bf Linear equations.} In this section we shall contruct soliton-antisoliton solutions. For this we choose the dressing factor $\chi$ to be
\begin{equation}
 \chi=1+\frac{2 i}{\zeta-i}\tilde{P},
\end{equation}
since if one uses $\text{Im}(\mu_k)<0$ for solitons then antisolitons correspond to $\mu_k$'s with $\text{Im}(\mu_k)>0$ (see e.g. \cite{Ioannidou:1996bb}). Once again, the reality condition (\ref{rc}) teaches us that
\begin{equation}
 \tilde{P}^\dagger=\tilde{P}\qquad\text{and}\qquad\tilde{P}^2=\tilde{P},
\end{equation}
such that $\tilde{P}$ can be considered to be a Hermitian projector $\tilde{P}=\tilde{T}(\tilde{T}^\dagger\tilde{T})^{-1}\tilde{T}^\dagger$. Since $\tilde{\psi}$ is again a solution of (\ref{l-pdes}) with some $\tilde{A}$ and $\tilde{B}$, we discover
\begin{eqnarray}
 \tilde{A}(t,x,y) &=& -\tilde{\psi}(t,x,y,\zeta)(\zeta\partial_x-\partial_u)
                                [\tilde{\psi}(t,x,y,\bar{\zeta})]^\dagger\notag\\
                  &=& \biggl(1+\frac{2i}{\zeta-i}\tilde{P}\biggr)\biggl(A-(\zeta\partial_x-\partial_u)\biggr)
                      \biggl(1-\frac{2i}{\zeta+i}\tilde{P}\biggr),\\
 \tilde{B}(t,x,y) &=& -\tilde{\psi}(t,x,y,\zeta)(\zeta\partial_v-\partial_x)
                                [\tilde{\psi}(t,x,y,\bar{\zeta})]^\dagger\notag\\
                  &=& \biggl(1+\frac{2i}{\zeta-i}\tilde{P}\biggr)\biggl(B-(\zeta\partial_v-\partial_x)\biggr)
                      \biggl(1-\frac{2i}{\zeta+i}\tilde{P}\biggr).
\end{eqnarray}
The poles at $\zeta=\pm i$ on the r.h.s. have to be removable since $\tilde{A}$ and $\tilde{B}$ do not depend on $\zeta$. Putting to zero the corresponding residues, we obtain the following pair of equations
\begin{equation}\label{sol-asol1}
 (1-\tilde{P})\left\{\partial_z\tilde{P}-(\partial_zP)\tilde{P}\right\} = 0\qquad\text{and}\qquad
 (1-\tilde{P})\left\{\frac{i}{2}\partial_t\tilde{P}+(\partial_{\bar{z}}P)\tilde{P}\right\}=0.
\end{equation}
Using the identities
\begin{equation}
 (1-\tilde{P})\tilde{P}=0\qquad\text{and}\qquad (1-\tilde{P})\tilde{T}=0
\end{equation}
the equations (\ref{sol-asol1}) reduce to the form
\begin{equation}\label{sol-asol2}
 (1-\tilde{P})\left\{a^\dagger\tilde{T}-[a^\dagger,P]\tilde{T}\right\}
        \frac{1}{\tilde{T}^\dagger\tilde{T}}\tilde{T}^\dagger = 0\quad\text{and}\quad
 (1-\tilde{P})\left\{\partial_t\tilde{T}-i\gamma[a,P]\tilde{T}\right\}
        \frac{1}{\tilde{T}^\dagger\tilde{T}}\tilde{T}^\dagger=0,
\end{equation}
with $\gamma=\sqrt{2/\theta}$. Obviously, these equations are solved by
\begin{equation}\label{sol-asol3}
 a^\dagger\tilde{T}-[a^\dagger,P]\tilde{T}=\tilde{T}Z_1 \qquad\text{and}\qquad
 \partial_t\tilde{T}-i\gamma[a,P]\tilde{T}=\tilde{T}Z_2,
\end{equation}
where $Z_{1,2}(t,a,a^\dagger)$  are some operator-valued functions.\\

\noindent {\bf Non-Abelian soliton-antisoliton configurations.} Again, we build the dressing factor $\chi$ with a matrix $\tilde{T}^\dagger=(\bar{u}_1,\ldots,\bar{u}_n)$ and choose the ansatz for $P$ to be of the form (\ref{ansatz-u(n)}). This implies for the first equation of (\ref{sol-asol3}) 
\begin{eqnarray}
 \bar{z}u_1-[\bar{z},K]\biggl(u_1+\sum_{i=1}^{n-1}\bar{z}^{m_i}u_{i+1}\biggr) &=& u_1Z_1,\\
 \bar{z}u_{j+1}-[\bar{z},z^{m_j}K]\biggl(u_1+\sum_{i=1}^{n-1}\bar{z}^{m_i}u_{i+1}\biggr) &=& u_{j+1}Z_1,
\end{eqnarray}
after a rescaling of $Z_1$. Multiplying the second of these equations by $\bar{z}^{m_j}$, summing over $j$ and adding to the first one yields
\begin{equation}
 \bar{z}\biggl(u_1+\sum_{j=1}^{n-1}\bar{z}^{m_j}u_{j+1}\biggr)
           =\biggr(u_1+\sum_{j=1}^{n-1}\bar{z}^{m_j}u_{j+1}\biggr)Z_1.
\end{equation}
Let us therefore choose $Z_1$ to be $\bar{z}$, i.e.
\begin{equation}
 \biggl[\bar{z},u_1+\sum_{j=1}^{n-1}\bar{z}^{m_j}u_{j+1}\biggr]=0\qquad\Rightarrow\qquad 
        u_1=-\sum_{j=1}^{n-1}\bar{z}^{m_j}u_{j+1}+f(t,\bar{z}),
\end{equation}
where $f(t,\bar{z})$ is arbitrary, but independent of $z$. Putting $Z_2$ identically to zero, the second equation of (\ref{sol-asol3}) reduces to
\begin{equation} 
 i\theta\partial_t\tilde{T}+[z,P]\tilde{T}=0
\end{equation}
and hence,
\begin{equation}
 i\theta\partial_tu_1=[K,z]f-2\theta K\sum_{j=1}^{n-1}m_j\bar{z}^{m_j-1}u_{j+1}.
\end{equation}
Defining 
\begin{equation}
 u_{i+1}=-1+z^{m_i}Kf
\end{equation}
it is not hard to verify that 
\begin{equation}
 f(t,\bar{z})=-2i\biggl(t\sum_{j=1}^{n-1}m_j\bar{z}^{m_j-1}+h(\bar{z})\biggr)
\end{equation}
solves our equations consistently with $h$ some operator-valued holomorphic function of $\bar{z}$. Hence,
\begin{equation}
 \tilde{T}=\begin{pmatrix} \sum_{i=1}^{n-1}\bar{z}^{m_i} \\ -1 \\ \vdots \\ -1 \end{pmatrix} +
           \begin{pmatrix} 1 \\ z^{m_1} \\ \vdots \\ z^{m_{n-1}} \end{pmatrix}\,
           \frac{-2i}{1+\sum_{i=1}^{n-1}\bar{z}^{m_i}z^{m_i}}\,
           \biggl(t\sum_{j=1}^{n-1}m_j\bar{z}^{m_j-1}+h(\bar{z})\biggr).
\end{equation}

\noindent {\bf Examples.} Let us now spezialize to the $n=2$ case. Therefore our solution reduces to
\begin{equation}\label{sol-asol-sol-u(2)}
 \tilde{T}=\begin{pmatrix} \bar{z}^m \\ -1 \end{pmatrix}+\begin{pmatrix} 1 \\ z^m \end{pmatrix}
           \frac{-2i}{1+\bar{z}^mz^m}\{m\bar{z}^{m-1}t+h(\bar{z})\},
\end{equation}
with $m:=m_1$. Depending on the explicit form of $h(\bar{z})$, the solutions (\ref{sol-asol-sol-u(2)}) describe different kinds of soliton-antisoliton configurations. Let us take, for instance, $h(\bar{z})=\frac{1}{2}\bar{z}^{2m+n}$ with $n\in\mathbb{N}\cup\{0\}$. This leads to a configuration consisting of $2m+n$ lumps. Obviously, the solution (\ref{sol-asol-sol-u(2)}) coincides with Ioannidou's given in \cite{Ioannidou:1996bb} in the commutative limit. 

To visualize the field configurations we perform the inverse Weyl transform (\ref{inv-weyl}) and find
\begin{equation}
 T\mapsto T_{\star}=\begin{pmatrix} 1\\ z^m \end{pmatrix} \quad\text{and}\quad
 \tilde{T}\mapsto\tilde{T}_{\star}=\begin{pmatrix} \bar{z}^m \\ -1 \end{pmatrix}
                                   +\begin{pmatrix} 1 \\ z^m \end{pmatrix}\star
           \frac{-2i}{1+\bar{z}^m\star z^m}\star\{m\bar{z}^{m-1}t+h_{\star}(\bar{z})\}.
\end{equation}
Hence, we know the field $\Phi_{\star}$ and can calculate its energy density $\mathcal{E}_{\star}$. Furthermore, for large distances $r^2=z\bar{z}$ the field $\Phi_{\star}$ and the energy density $\mathcal{E}_{\star}$ will appoach their commutative limits $\Phi$ and $\mathcal{E}$, such that the analysis given in \cite{Ioannidou:1996bb} can be applied in the same way.

\begin{figure}[ht]
 \begin{minipage}[h]{2.5in}
  \begin{center}
   \includegraphics[height=1.5in,width=2.5in]{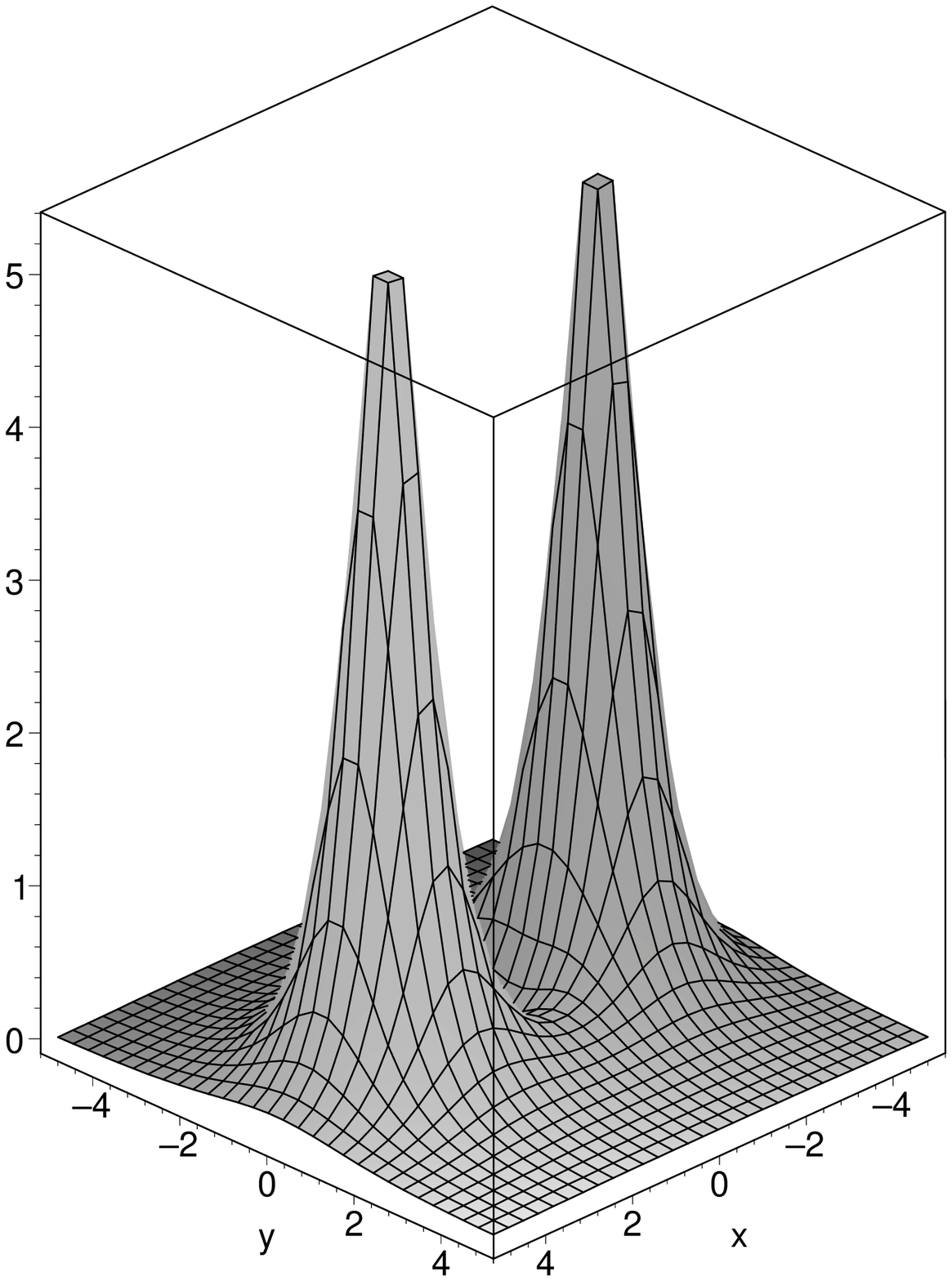}
   \caption{\small Energy density for one $U(2)$ soliton and one antisoliton, 
                   showing right angle scattering at time $t=-5$ for $\theta\ll 1$.}
   \label{scatsol-soas1}
  \end{center}
 \end{minipage}
 \hfill
 \begin{minipage}[h]{2.5in}
  \begin{center}
   \includegraphics[height=1.5in, width=2.5in]{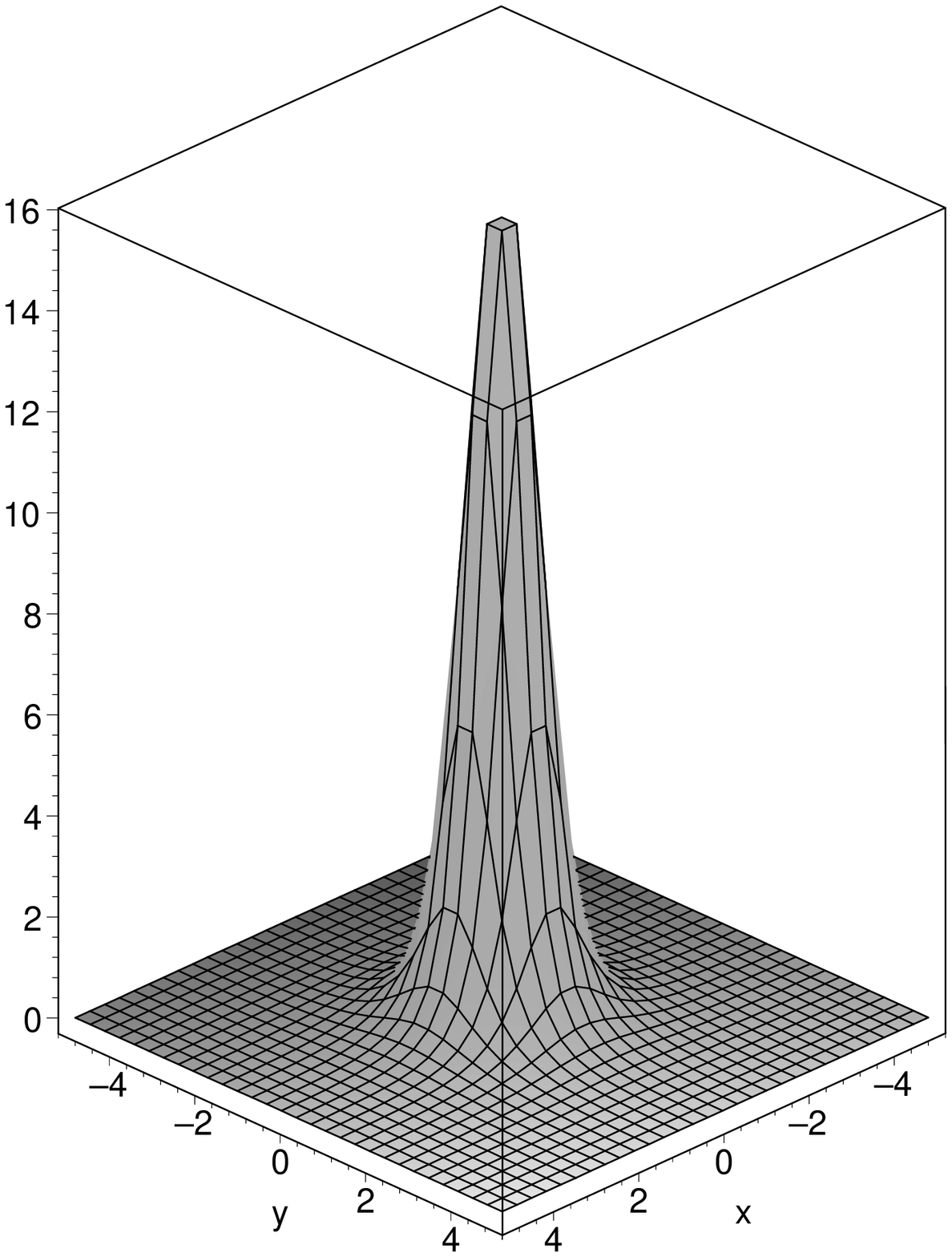}
   \caption{\small Energy density for one $U(2)$ soliton and one antisoliton, 
                   showing right angle scattering at time $t=0$ for $\theta\ll 1$.}
   \label{scatsol-soas2}
  \end{center}
 \end{minipage}
\end{figure}
\begin{figure}[ht]
 \begin{center}
   \includegraphics[height=1.5in,width=2.5in]{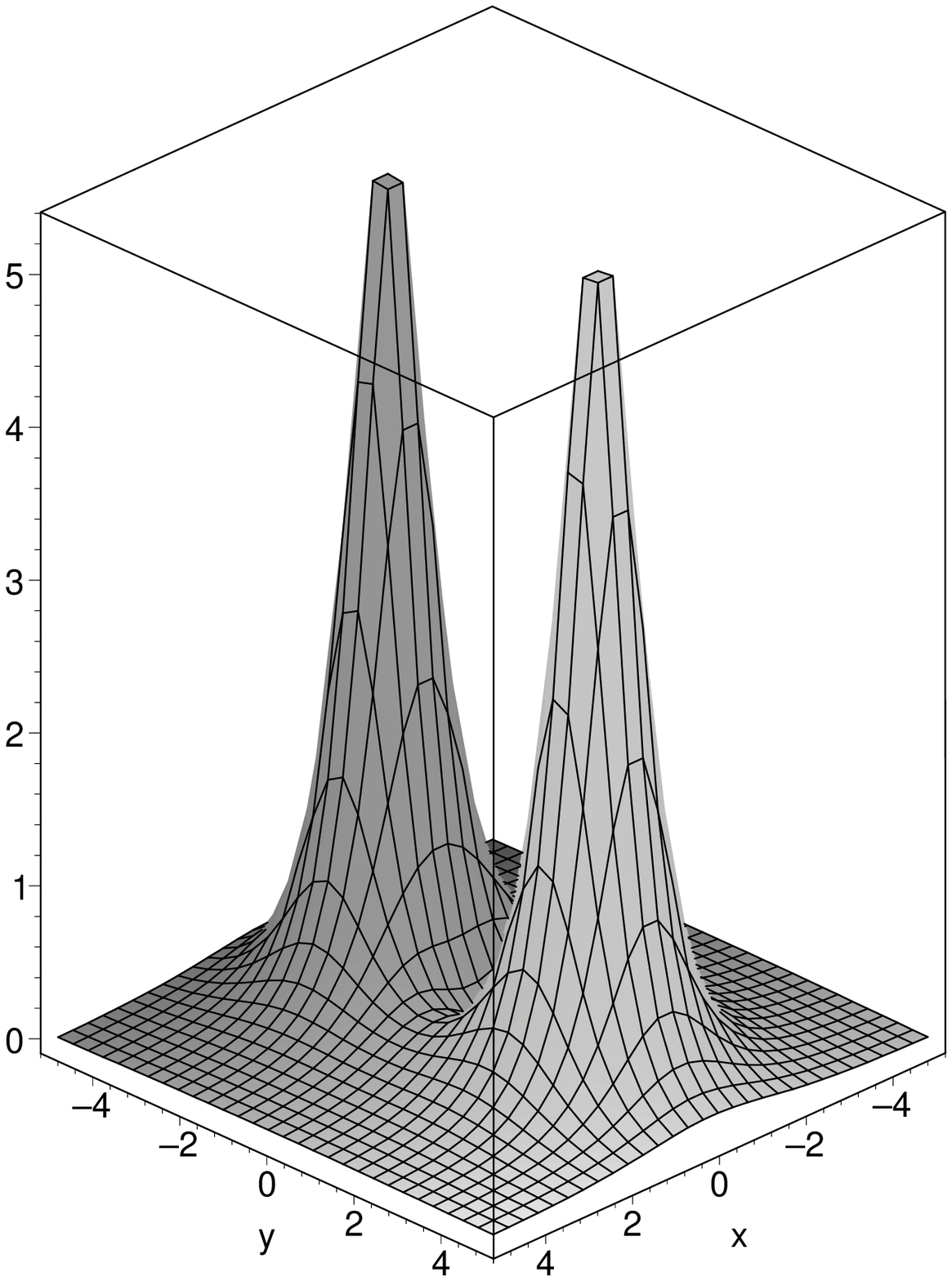}
   \caption{\small Energy density for one $U(2)$ soliton and one antisoliton, 
                   showing right angle scattering at time $t=5$ for $\theta\ll 1$.}
   \label{scatsol-soas3}
  \end{center}
\end{figure}

The simplest scattering case occurs for $m=1$ and $n=0$, i.e. $h(\bar{z})=\frac{1}{2}\bar{z}^2$. After a few lines of algebra one finds that for large $r$ the energy density is given by
\begin{equation} 
 \mathcal{E}_{\star}=16\frac{2r^4+4r^2+4t^2(1+2r^2)-4t(x^2-y^2)+1}{[2r^4+2r^2+4t(x^2-y^2)+4t^2+1]^2}
                     [1+O(\theta/r^2)],
\end{equation}
i.e. it peaks near the locus $\bar{z}^2+2t=0$. To be more precise, by varying $t$ we see the two lumps accelerating symmetrically towards each other along the $x$-axis, interacting at the origin (near $t=0$) and decelerating to infinity along the $y$-axis. This configuration has the topological charge $q=0$ since as argued above it coincides with Ioannidou's solution \cite{Ioannidou:1996bb} in the limit $\theta\to 0$ and the topological charge is independent of $\theta$. Thus, a head-on collision of one soliton and one antisoliton results in a $90^{\text{o}}$ angle scattering. Figures \ref{scatsol-soas1} - \ref{scatsol-soas3} show this case for small $\theta$.

\begin{figure}[ht]
 \begin{minipage}[h]{2.5in}
  \begin{center}
   \includegraphics[height=1.5in,width=2.5in]{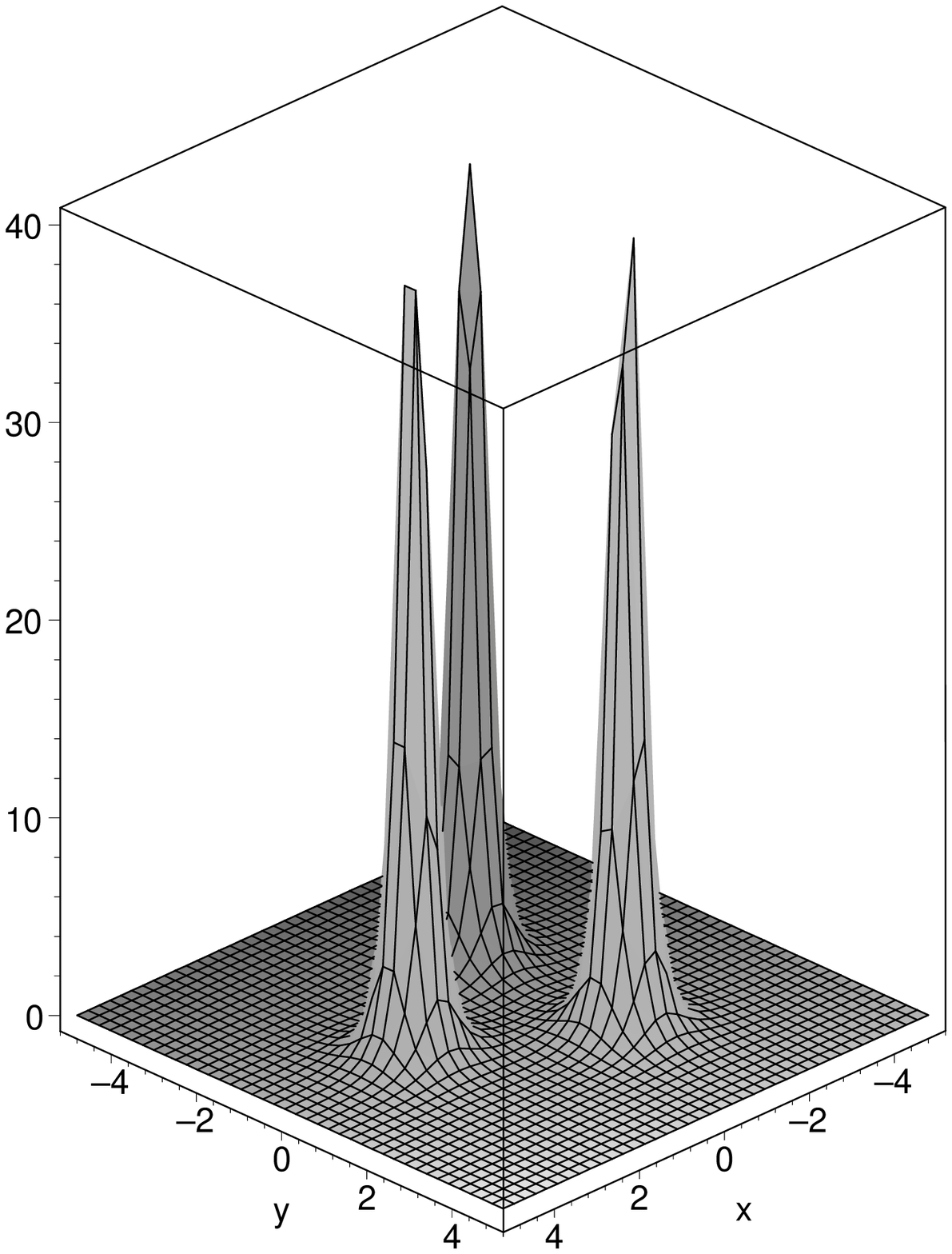}
   \caption{\small Energy density for one $U(2)$ soliton and two antisolitons, 
                   showing $60^{\text{o}}$ angle scattering at time $t=-5$ for $\theta\ll 1$.}
   \label{scatsol-so2as1}
  \end{center}
 \end{minipage}
 \hfill
 \begin{minipage}[h]{2.5in}
  \begin{center}
   \includegraphics[height=1.5in, width=2.5in]{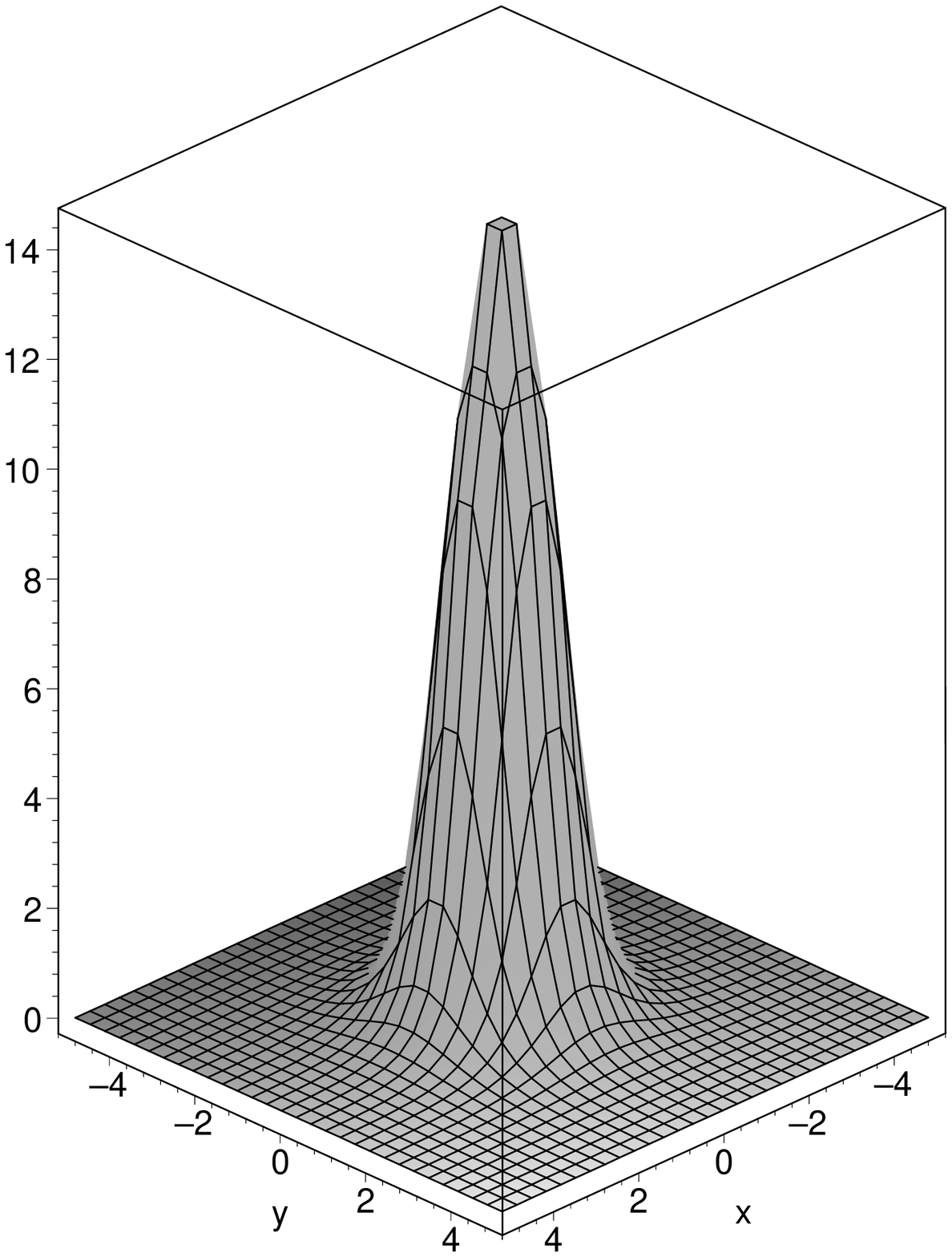}
   \caption{\small Energy density for one $U(2)$ soliton and two antisolitons, 
                   showing $60^{\text{o}}$ angle scattering at time $t=0$ for $\theta\ll 1$.}
   \label{scatsol-so2as2}
  \end{center}
 \end{minipage}
\end{figure}

\begin{figure}[ht]
 \begin{center}
   \includegraphics[height=1.5in,width=2.5in]{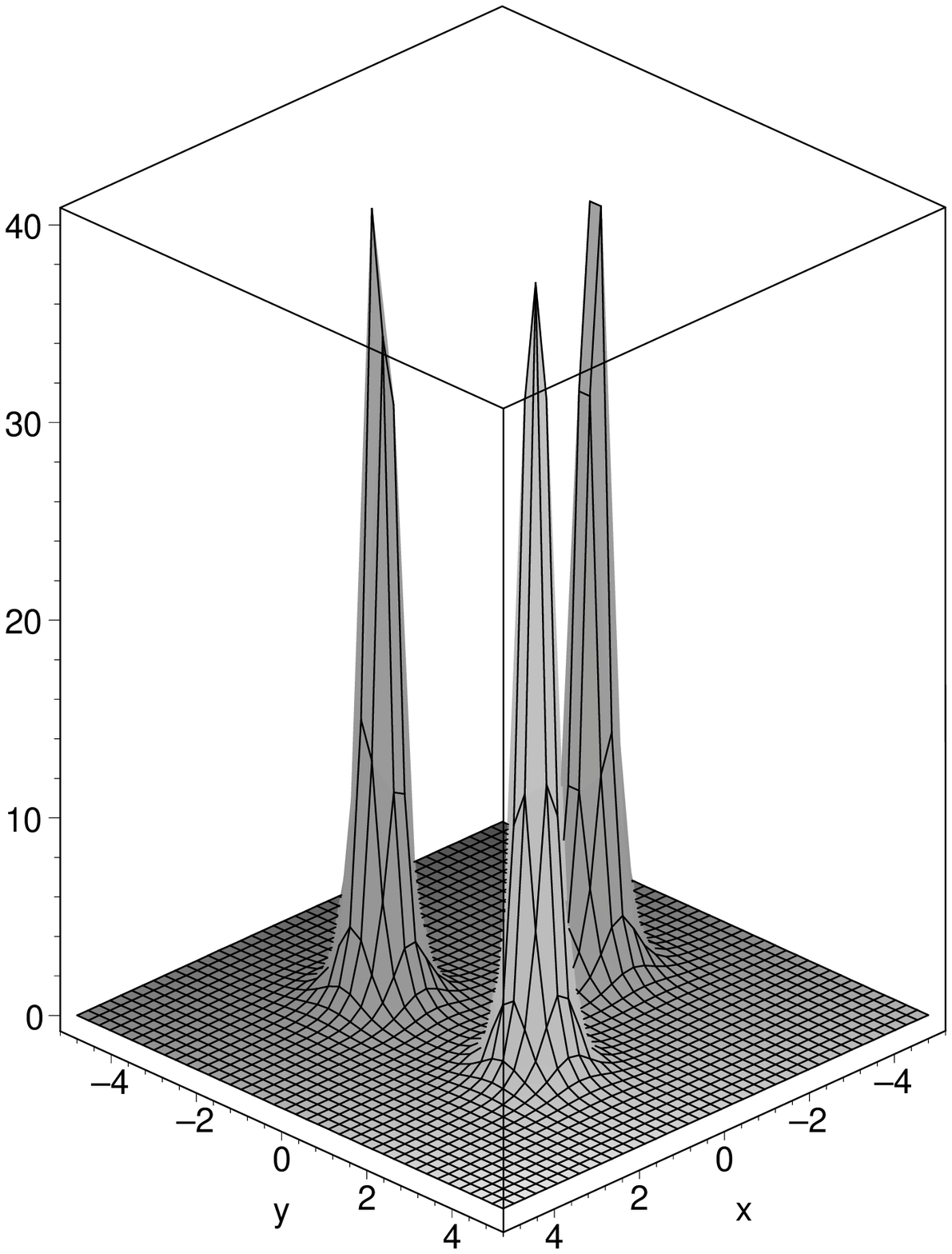}
   \caption{\small Energy density for one $U(2)$ soliton and two antisolitons, 
                   showing $60^{\text{o}}$ angle scattering at time $t=5$ for $\theta\ll 1$.}
   \label{scatsol-so2as3}
  \end{center}
\end{figure}

To exhibit a scattering configuration with one soliton and two antisolitons we take $m=1$ and $n=1$, i.e. $h(\bar{z})=\frac{1}{2}\bar{z}^3$ (i.e. $q=-1$). In that case the energy density turns out to be
\begin{equation}
 \begin{split}
 \mathcal{E}_{\star} = 
  &8[r^8+8r^6+11r^4+4r^2-8x^5t+16ty^2(x^3+t)+8t^2+48xy^2t-16x^2t(x-t)\\
  &+24xty^4+2]/[r^6+r^4+2r^2+4t^2+4tx^3-12xy^2t+1]^2[1+O(\theta/r^2)].
 \end{split}
\end{equation}
Finding its extremas we realize that a head-on collision of one soliton and two antisolitons results in a $60^{\text{o}}$ angle scattering, as can be seen from figures \ref{scatsol-so2as1} - \ref{scatsol-so2as3}.


Finally, we note, that in both cases the noncommutativity is only felt inside a disc of radius $R\sim\sqrt{\theta}$, such that for large distances their behavior agrees with their commutative pendants.


\section{Concluding remarks}

In this paper we have applied the method of dressing transformations to a noncommutative modified $U(n)$ sigma model in order to generate non-Abelian multi-soliton and soliton-antisoliton solutions. However, a natural question does arise regarding the construction of Abelian soliton-antisoliton configurations. In \cite{Lechtenfeld:2001gf} it was shown, that Abelian multi-soliton configurations do exist and actually they were constructed. Hence, pure Abelian antisolitions exist as well, since we have an ambiguity in what we are calling soliton and what antisoliton, i.e. the sign of the definition of the topological charge is rather a matter of taste. However, no Abelian scattering solutions of the modified sigma model in 2+1 dimensions have been found so far, rather ring-like structures with solitons (or antisolitons) sitting at the origin and changing their shape with time.

Considering the Abelian soliton-antisoliton case lead to even more negative results in such a way, that using the techniques discussed above we did not obtain any stable configuration. It might be possible that Abelian solutions do exist considering higher order poles in (\ref{ansatz-dressing}).

Finally, noncommutative multi-instanton solutions in noncommutative Yang-Mills theory were constructed via the ADHM approach (see e.g. \cite{Nekrasov:1998ss} and \cite{Chu:2001cx}) and using the twistor approach (see \cite{Lechtenfeld:2001ie}). They can also be obtained by employing the dressing method, as it was done in the commutative setup in \cite{Belavin:1978pa} and \cite{Forgacs:1981su}. Thus, it would be interesting to use the dressing method to construct noncommutative multi-instanton solutions and to compare the results to those obtained by the ADHM and by the twistor approach. It is also interesting to apply the dressing approach to the construction of soliton-like solutions in Berkovits' string field theory \cite{Berkovits:1995ab} as was initiated in \cite{Lechtenfeld:2000qj},\cite{Lechtenfeld:2002cu}.

\section*{Acknowledgements}

I would like to thank Alexander Popov for helpful comments. Further I am grateful to the Center for Geometry and Theoretical Physics, Duke University and to the Studienstiftung des deutschen Volkes for financial support. This work was done within the framework of the DFG priority program in string theory.

\end{document}